\documentclass[12pt]{article}

\usepackage{graphicx}

\begin{document}

\title{Decoherence of the Kondo Singlet Caused by Phase-sensitive Detection}

\author{Ming-Lun Chen \\$Department of Physics, Jinggangshan University,
Ji'an 343009, China$\\}

\maketitle

\begin{abstract}
We investigate the dephasing effect of the Kondo singlet in an
Aharonov-Bohm interferometer with a quantum dot coupling to left and
right electrodes. By employing the cluster expansions, the equations
of motion of Green functions are transformed into the corresponding
equation of motion of connected Green functions, which contains the
correlation of two conduction electrons beyond the Lacroix
approximation. With the method we show that the Kondo resonance is
suppressed by phase-sensitive detection of Aharonov-Bohm
interferometer. Our numerical results have provided a qualitative
explanation with the anomalous features observed in a recent
experiment by Avinun-Kalish \emph{et al}. [Phys. Rev. Lett.
\textbf{92}, 156801 (2004)].
\end{abstract}

{\textbf{PACS}: 73.23.-b, 73.63.-b, 72.15.Qm, 75.20.Hr}

Controlled dephasing experiments in mesoscopic devices provide an
excellent playground for probing the nature of phase coherence
transport and studying the wave-particle duality in quantum
mechanics. In the devices, coherence of the quantum dots (QD) was
monitored by an Aharonov-Bohm interferometer (ABI), and decoherence
was induced by a quantum point contact (QPC). Initially such
experiments were performed in the mesoscopic structures based on QD
in the Coulomb blockade regime \cite{1,2,3}. Then this kind of
experiment was extended to the Kondo regime of QD. In the Kondo
regime, a Kondo singlet is formed between the localized spin in a QD
and electrons in the electrodes. It was shown that the existence of
the QPC plays a role of path-sensitive detector and raises
significant suppression of the Kondo resonance \cite{4}. However,
properties of the suppression were strongly different from the
theoretical prediction of Ref. [5]. The most significant deviation
from the theory is that the measured suppression strength of the
Kondo resonance is larger 30 times than expected.

Recently, to explain the anomalous features, a theory was proposed
by K. Kang \cite{6}, in which K. Kang thought that the
phase-sensitive detection of the QPC is also an important component
for the decoherence of the Kondo singlet. We point out that this
treatment is incomplete, because the phase-sensitive detection of
the QD is performed mainly by the ABI and not by the QPC. Therefore
phase-sensitive detection of the ABI should also be taken into
account. One way to prove our proposal is to throw off the QPC from
the controlled dephasing devices and only to check the influence of
phase-sensitive detection of the ABI, then the controlled depasing
device becomes an Aharonov-Bohm interferometer with a quantum dot
coupling to left and right electrodes, which is designed first by
Yacoby \cite{7} to measure the phase sensitivity of a QD. It is just
the model that we intend to investigate.

In this Letter, we provide a qualitative explanation with the
anomalous features observed in a recent dephasing experiment by
Avinun-Kalish \emph{et al}. By employing the cluster expansions, the
equations of motion (EOM) of Green's functions are transformed into
the corresponding EOM of connected Green's functions, which contains
the correlation of two conduction electrons beyond the Lacroix
approximation. With the method we investigate the Kondo effect in an
Aharonov-Bohm interferometer with a quantum dot coupling to left and
right electrodes. The differential conductance of the system are
calculated to show that the Kondo resonance is suppressed by
phase-sensitive detection of the ABI. Our numerical results have
shown that the theory of K. Kang is incomplete and the
phase-sensitive detection of the ABI should also be taken account.

An Aharonov-Bohm interferometer with a quantum dot coupling to left
and right electrodes can be modeled by the following Hamiltonian:
\begin{eqnarray}
H &=&\sum\limits_{\alpha k\sigma }\varepsilon _{\alpha k}C_{\alpha
k\sigma }^{\dagger }C_{\alpha k\sigma }+\sum\limits_{\sigma
}\varepsilon _{d\sigma }d_{\sigma }^{\dagger }d_{\sigma
}+\frac{U}{2}\sum\limits_{\sigma }n_{\sigma }n_{\bar{\sigma}}
+\sum\limits_{\alpha k\sigma }(V_{\alpha }d_{\sigma }^{\dagger
}C_{\alpha k\sigma }+V_{\alpha }^{\ast }C_{\alpha k\sigma }^{\dagger
}d_{\sigma })
\nonumber\\
&&+\sum\limits_{kk^{^{\prime }}\sigma }(T_{LR}C_{Lk\sigma }^{\dagger
}C_{Rk^{^{\prime }}\sigma }+T_{LR}^{\ast }C_{Rk^{^{\prime }}\sigma
}^{\dagger }C_{Lk\sigma }),
\end{eqnarray}
where $\alpha=L, R$ denotes the left or right electrode, and
$\sigma=\uparrow, \downarrow$ denotes spin up or down.The first term
describes electrons in the left and the right electrodes, and the
second one describes electrons of the quantum dot. The third one
corresponds to the on-site Coulomb interactions, and $U$ is the
on-site Coulomb repulsion. The fourth one describes the tunneling
through the quantum dot, and $V_{\alpha}$ represents the $s-d$
hybridization. The last one describes the tunneling of electrons
between two electrodes via the direct channel, and $T_{LR}$ is the
direct electron transmission. The Aharonov-Bohm phase
$\phi=2\pi\Phi\times e/hc$ is included in the tunneling matrices as
$V_{L}T_{LR}V_{R}=|V_{L}T_{LR}V_{R}|e^{i\phi}$. $\Phi$ is the
magnetic flux enclosed in the Aharonov-Bohm ring.

Following Zubarev \cite{8}, the retarded Green's function
$\langle\langle A(t);B(t')\rangle\rangle$ is defined as
\begin{equation}
\langle\langle
A(t);B(t')\rangle\rangle=-i\theta(t-t')\langle[A(t),B(t')]_\pm\rangle.
\end{equation}
By the Fourier transformation of the time variable, the retarded
Green's function satisfies the following equation:
\begin{equation}
\omega\langle\langle
A;B\rangle\rangle=\langle[A,B]_\pm\rangle+\langle\langle[A,H]_-;B\rangle\rangle.
\end{equation}
Eq.(3) is named as the equation of motion(EOM) of Green's function
for the Hamiltonian(1), which can be expressed specifically as
follows:
\begin{eqnarray}
\omega\langle\langle d_\sigma;d_\sigma^\dag\rangle\rangle
=1+\varepsilon_{d\sigma}\langle\langle
d_\sigma;d_\sigma^\dag\rangle\rangle +\sum_{\alpha k}
V_\alpha\langle\langle C_{\alpha
k\sigma};d_\sigma^\dag\rangle\rangle+U\langle\langle
n_{\bar{\sigma}} d_\sigma;d_\sigma^\dag\rangle\rangle.
\end{eqnarray}
$G_{d\sigma}(\omega)=\langle\langle
d_\sigma;d_\sigma^\dag\rangle\rangle$ is the Green's function of the
QD, and for the high-order Green's function
\begin{eqnarray}
\omega\langle\langle n_{\bar{\sigma}}d_{\sigma };d_{\sigma
}^{\dagger }\rangle\rangle&=&n_{\bar{\sigma}}+(\varepsilon _{d\sigma
}+U)\langle\langle n_{\bar{\sigma}}d_{\sigma };d_{\sigma }^{\dagger
}\rangle\rangle+\sum\limits_{\alpha k}V_{\alpha }\langle \langle
n_{\bar{\sigma}}C_{\alpha k\sigma };d_{\sigma }^{\dagger
}\rangle\rangle+\sum\limits_{\alpha k}V_{\alpha }\langle\langle
d_{\bar{\sigma}}^{\dagger }C_{\alpha k\bar{\sigma}}d_{\sigma
};d_{\sigma }^{\dagger }\rangle\rangle
\nonumber\\
&&-\sum\limits_{\alpha k}V_{\alpha }^{\ast
}\langle\langle C_{\alpha k\bar{\sigma}}^{\dagger }d_{\bar{\sigma}%
}d_{\sigma };d_{\sigma }^{\dagger }\rangle\rangle,
\end{eqnarray}
instead of employing directly Lacroix decoupling scheme \cite{9, 10},
we make use of a cluster expansions to separate the connected part
of the Green's function. As an example, the high-order Green's
function $\langle\langle n_{\bar{\sigma}}d_{\sigma};d_{\sigma
}^{\dagger }\rangle\rangle$ is expressed as follows:
\begin{eqnarray}
\langle\langle n_{\bar{\sigma}}d_{\sigma };d_{\sigma }^{\dagger
}\rangle\rangle=\langle n_{\bar{\sigma}}\rangle\langle\langle
d_{\sigma };d_{\sigma }^{\dagger }\rangle\rangle+\langle\langle
n_{\bar{\sigma}}d_{\sigma };d_{\sigma }^{\dagger
}\rangle\rangle_c\nonumber ,
\end{eqnarray}
where $\langle\langle \cdot \cdot \cdot\rangle\rangle_c$ represents
a connected Green's function and it is straightforward to derive the
EOM. We write down the EOM of the connected Green's function
$\langle \langle n_{\bar{\sigma}}d_{\sigma };d_{\sigma }^{\dagger
}\rangle \rangle _{c}$ as follows:
\begin{eqnarray}
(\omega-\varepsilon _{d\sigma }&-&U(1-n_{\bar{\sigma}}))\langle
\langle n_{\bar{\sigma}}d_{\sigma };d_{\sigma }^{\dagger }\rangle
\rangle _{c}=Un_{\bar{\sigma}}(1-n_{\bar{\sigma}})\langle \langle
d_{\sigma };d_{\sigma }^{\dagger }\rangle\rangle
+\sum\limits_{\alpha k}V_{\alpha }\langle\langle n_{\bar{\sigma}
}C_{\alpha k\sigma };d_{\sigma }^{\dagger }\rangle\rangle _{c}
\nonumber\\
&+&\sum\limits_{\alpha k}V_{\alpha }\langle\langle d_{\bar{
\sigma}}^{\dagger }C_{\alpha k\bar{\sigma}}d_{\sigma };d_{\sigma
}^{\dagger }\rangle\rangle _{c}-\sum\limits_{\alpha k}V_{\alpha
}^{\ast }\langle\langle C_{\alpha k\bar{\sigma}}^{\dagger
}d_{\bar{\sigma}}d_{\sigma };d_{\sigma }^{\dagger }\rangle\rangle
_{c}.
\end{eqnarray}
It is not difficult to obtain the EOM of the other connected Green's
function such as $\langle\langle n_{\bar{\sigma}}C_{\alpha
k\sigma};d_{\sigma }^{\dagger }\rangle\rangle_{c}$, $\langle\langle
d_{\bar{\sigma}}^{\dagger }C_{\alpha k\bar{\sigma}}d_{\sigma
};d_{\sigma}^{\dagger }\rangle\rangle _{c}$, $\langle\langle C_{\alpha k\bar{\sigma}}^{\dagger }d_{\bar{\sigma}%
}d_{\sigma };d_{\sigma }^{\dagger }\rangle\rangle _{c}$, $
\langle\langle d_{\bar{\sigma}}^{\dagger }C_{\alpha ^{\prime
}k^{\prime }\bar{\sigma}}C_{\alpha k\sigma };d_{\sigma }^{\dagger
}\rangle\rangle _{c}$, $\langle\langle C_{\alpha ^{\prime }k^{\prime
}\bar{\sigma}}^{\dagger }d_{\bar{\sigma}}C_{\alpha k\sigma
};d_{\sigma }^{\dagger }\rangle\rangle _{c}$, and $\langle\langle C_{\alpha ^{\prime }k^{\prime }\bar{\sigma}%
}^{\dagger }C_{\alpha k\bar{\sigma}}d_{\sigma };d_{\sigma }^{\dagger
}\rangle\rangle _{c}$. In order to truncate the EOM, if let
$\langle\langle d_{\bar{\sigma}}^{\dagger }C_{\alpha ^{\prime
}k^{\prime }\bar{\sigma}}C_{\alpha k\sigma };d_{\sigma }^{\dagger
}\rangle\rangle _{c}$, $\langle\langle C_{\alpha ^{\prime }k^{\prime
}\bar{\sigma}}^{\dagger }d_{\bar{\sigma}}C_{\alpha k\sigma
};d_{\sigma }^{\dagger }\rangle\rangle _{c}$, and $\langle\langle C_{\alpha ^{\prime }k^{\prime }\bar{\sigma}%
}^{\dagger }C_{\alpha k\bar{\sigma}}d_{\sigma };d_{\sigma }^{\dagger
}\rangle\rangle _{c}$ involving with the correlation of two
conduction electrons to be zero, one reaches Lacroix approximation.

As a consequence of a coherent superposition of spin flip
cotunneling events, the forming of the Kondo resonance are
inevitably involved with the correlation of two conduction
electrons. Therefore the Lacroix approximation must be improved in
order to discus the electron transport properties of the
single-impurity Anderson model. To surpass Lacroix approximation it
is necessary to consider the EOM of the connected Green's functions
of $\langle\langle d_{\bar{\sigma}}^{\dagger }C_{\alpha ^{\prime
}k^{\prime }\bar{\sigma}}C_{\alpha k\sigma };d_{\sigma }^{\dagger
}\rangle\rangle _{c}, \langle\langle C_{\alpha ^{\prime }k^{\prime
}\bar{\sigma}}^{\dagger }d_{\bar{\sigma}}C_{\alpha k\sigma
};d_{\sigma }^{\dagger }\rangle\rangle _{c}$, and $
\langle\langle C_{\alpha ^{\prime }k^{\prime }\bar{\sigma}%
}^{\dagger }C_{\alpha k\bar{\sigma}}d_{\sigma };d_{\sigma }^{\dagger
}\rangle\rangle _{c}$ involving the correlation of two conduction
electrons, and to assume higher-order correlation Green's functions
to be zero. After a lengthy but direct calculation, in the limit of
$U\rightarrow\infty$, $\langle\langle
d_\sigma;d_\sigma^\dag\rangle\rangle$ is obtained finally as
\begin{eqnarray}
G_{d\sigma }=\frac{1-n_{\bar{\sigma}}-\sum\limits_{\alpha
k}\frac{V_{\alpha }\langle d_{\bar{\sigma}}^{\dagger }C_{\alpha
k\bar{\sigma}}\rangle }{\omega -\varepsilon _{\alpha
k}}-\frac{V^2_{\alpha }}{n_{\bar{\sigma}}}\sum\limits_{\alpha
k}\frac{\langle d_{\bar{\sigma}}^{\dagger }C_{\alpha
k\bar{\sigma}}\rangle }{\omega -\varepsilon _{\alpha
k}}\sum\limits_{\alpha k'}\frac{\langle d_{\bar{\sigma}}^{\dagger
}C_{\alpha k'\bar{\sigma}}\rangle }{\omega -\varepsilon _{\alpha
k'}}}{\omega -\varepsilon _{d\sigma }-\sum\limits_{\alpha k}\frac{
V^2_{\alpha }}{\omega -\varepsilon _{\alpha k}} +\sum\limits_{\alpha
k}\frac{V^2_{\alpha }}{\omega -\varepsilon _{\alpha
k}}\sum\limits_{\alpha k}\frac{V_{\alpha }\langle
d_{\bar{\sigma}}^{\dagger }C_{\alpha k\bar{\sigma} }\rangle }{\omega
-\varepsilon _{\alpha k}}-\sum\limits_{\alpha k}\sum\limits_{\alpha
^{\prime }k^{\prime }}\frac{ V_{\alpha }V_{\alpha ^{\prime }}^{\ast
}\langle C_{\alpha ^{\prime }k^{\prime }\bar{\sigma}}^{\dagger
}C_{\alpha k\bar{\sigma}}\rangle }{\omega -\varepsilon _{\alpha
k}}+\sum\limits_{\alpha k}\frac{ V_{\alpha }T_{LR}^{\ast }\langle
d_{\bar{\sigma}}^{\dagger }C_{\alpha k\bar{\sigma}}\rangle}{\omega
-\varepsilon _{\alpha k}}+\delta}.
\end{eqnarray}
where
\begin{eqnarray}
\delta &=& \frac{V^2_{\alpha }}{n_{\bar{\sigma}}}\sum\limits_{\alpha
kk'}\frac{\langle
d_{\sigma}^{\dagger}d_{\bar{\sigma}}^{\dagger}C_{\alpha
k'\bar{\sigma}}C_{\alpha k\sigma}\rangle_c-\langle
d_{\sigma}^{\dagger}C_{\alpha k'\bar{\sigma}}^{\dagger
}d_{\bar{\sigma}}C_{\alpha k\sigma }\rangle_c}{\omega -\varepsilon
_{\alpha k}}+\frac{V^2_{\alpha
}}{n_{\bar{\sigma}}}\sum\limits_{\alpha kk'}\frac{\langle
\hat{n}_{\bar{\sigma}}C_{\alpha k'\bar{\sigma}}^{\dagger }C_{\alpha
k\bar{\sigma}}\rangle_c-\langle d_{\bar{\sigma}}^{\dagger}C_{\alpha
k\bar{\sigma}}\rangle \langle C_{\alpha k'\bar{\sigma}}^{\dagger
}d_{\bar{\sigma}}\rangle }{\omega -\varepsilon
_{\alpha k}}\nonumber\\
&+&\frac{V^2_{\alpha }}{n_{\bar{\sigma}}}\sum\limits_{\alpha
kk'}\frac{\langle
d_{\sigma}^{\dagger}d_{\bar{\sigma}}^{\dagger}C_{\alpha
k'\bar{\sigma}}C_{\alpha k\sigma}\rangle_c-\langle
d_{\bar{\sigma}}^{\dagger}C_{\alpha k\bar{\sigma}}\rangle \langle
d_{\sigma}^{\dagger }C_{\alpha k'\sigma}\rangle }{\omega
-\varepsilon _{\alpha k}}-\frac{V^2_{\alpha
}T_{LR}}{n_{\bar{\sigma}}}\sum\limits_{\alpha kk'}\frac{\langle
d_{\bar{\sigma}}^{\dagger}C_{\alpha k\bar{\sigma}}\rangle \langle
C_{\alpha k"\sigma}^{\dagger }C_{\alpha k'\sigma}\rangle+\langle
\hat{n}_{\sigma}d_{\bar{\sigma}}^{\dagger}C_{\alpha
k'\bar{\sigma}}\rangle_c}{(\omega -\varepsilon _{\alpha k})(\omega
-\varepsilon
_{\alpha k'})}\nonumber\\
\end{eqnarray}
and the average functions $\langle d_{\bar{\sigma}}^{\dagger }C_{\alpha k\bar{\sigma}%
}\rangle $ and $\langle C_{\alpha ^{\prime }k^{\prime
}\bar{\sigma}}^{\dagger }C_{\alpha k\bar{\sigma}}\rangle$ can be
calculated by spectral theorem as follows
\begin{eqnarray}
\langle d_{\bar{\sigma}}^{\dagger }C_{\alpha k\bar{\sigma}}\rangle
=-\frac{1}{\pi}\int f(\omega )Im\langle \langle C_{\alpha
k\bar{\sigma}};d_{\bar{\sigma}}^{\dagger }\rangle \rangle,
\end{eqnarray}
\begin{eqnarray}
\langle C_{\alpha ^{\prime }k^{\prime }\bar{\sigma}}^{\dagger
}C_{\alpha k\bar{\sigma}}\rangle =-\frac{1}{\pi}\int f(\omega )
Im\langle \langle C_{\alpha k\bar{\sigma}};C_{\alpha ^{\prime
}k^{\prime }\bar{\sigma}}^{\dagger }\rangle \rangle,
\end{eqnarray}
where $f(\omega)=1/[exp((\omega-E_F)/T)+1]$ is the
Fermi-distribution function. The EOM of the corresponding Green's
functions read
\begin{eqnarray}
(\omega-\varepsilon_{\alpha k})\langle \langle C_{\alpha
k\bar{\sigma}};d_{\bar{\sigma}}^{\dagger }\rangle
\rangle=(V_{\alpha}-\frac{V_\alpha
T^*_{LR}}{\omega-\varepsilon_{\alpha k}})G_{d\bar{\sigma}}(\omega),
\end{eqnarray}
\begin{eqnarray}
(\omega-\varepsilon_{\alpha k})\langle \langle C_{\alpha
k\bar{\sigma}};C_{\alpha ^{\prime }k^{\prime }\bar{\sigma}}^{\dagger
}\rangle
\rangle=(\delta_{\alpha\alpha^{\prime}kk^\prime}-\frac{V_\alpha
V^*_{\alpha^\prime}}{\omega-\varepsilon_{\alpha k}} +\frac{V_\alpha
V^*_{\alpha^\prime}T_{LR}}{\omega-\varepsilon_{\alpha
k}})G_{d\bar{\sigma}}(\omega).
\end{eqnarray}
For the sake of simplicity, we have considered the nonmagnetic case,
i.e., $n_{d\downarrow}=n_{d\uparrow}=n/2$, which $n$ is total $d$
electron number
\begin{equation}
n=2n_{\bar{\sigma}}=\int
f(\omega^\prime)\rho(T,\omega^\prime)d\omega^\prime,
\end{equation}
where $\rho(T,\omega)=-(1/\pi)ImG_d(\omega)$ is the density of
states with finite temperature. Equations (7)-(12) constitute a
closed set of equations, which can be solved self-consistently and
numerically.
\begin{figure}[tbp]
\includegraphics[width=0.5\textwidth]{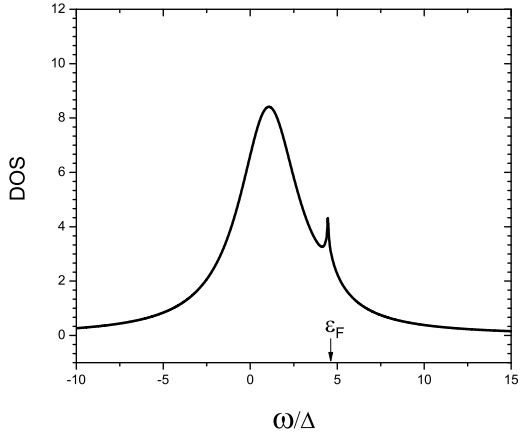}
\caption{DOS of quantum dot for $T_{LR}=0.01$. The position of the
Kondo resonance peak is labeled by the arrow.}
\end{figure}

In the following we calculate the density of states (DOS) of the QD
in the Kondo regime by
$\rho(T,\omega)=-(1/\pi)ImG_{d\sigma}(\omega)$. Because only the
conduction electrons near the Fermi level $\varepsilon_F$
participate in the transport current, the DOS for conduction
electrons is taken to be a constant $\rho(\varepsilon)=1/(2D)$ as
$-D<\varepsilon<D$, and the quantum dot level broadening is given by
$\Delta=\pi|V_\alpha|^2\rho(\varepsilon)$ \cite{11}. The Fermi
energy $\varepsilon_F$ (meV) is reference mark of the unit of energy
 \cite{7}. The parameters are considered in the following. The hopping matrix
element $V_{\alpha}$ between the quantum dot and the electrodes is
taken to be 0.1, and the tunneling matrix element $T_{LR}$ is taken
to be 0.01, which describes the electron direct transmission between
two electrodes via direct channel. The total number of the $d$
electron is taken to be 0.8, which determines self-consistently the
chemical potential and the numerical value can ensure that the
relative position of the level
$\Delta\varepsilon=\varepsilon_F-\varepsilon_d$ lies in the Kondo
regime. The half width $D$ is assumed to be 1, which defines the
energy scale. The $d$ electron level $\varepsilon_{d}$ is taken to
be 0. The magnetic flux $\Phi$ enclosed in the ring is taken to be
$0.5hc/e$ and the virtual dot level broadening $\Delta=0.01D$.
Figure 1 presents the DOS at the very low temperature
($T=10^{-5}\Delta$ which is lower than Kondo temperature
$T_{k}=(D\Delta)^{1/2}exp(\pi(\varepsilon_d-\varepsilon_F)/(2\Delta))$
\cite{12}). The Lorentzian resonance peak, which is the broadened
quantum dot level, is slightly shifted away from zero. At the Fermi
level a Kondo resonance peak is observed.

The current from the left to the right electrode can be calculated
from the time evolution of the occupation number of the left
electrode:
\begin{equation}
J_L(t)=-e\langle
\frac{dN_L}{dt}\rangle=-\frac{ie}{h}\langle[H,N_L]\rangle.
\end{equation}
where $N_L=\sum\limits_{k,\sigma}C^\dag_{kL,\sigma}C_{kL,\sigma}$,
Using the Green function of the Keldysh type \cite{13}
$G^<_{dLk\sigma}(t,t)$ and $G^<_{CLk\sigma,CRk'\sigma}(t,t)$
corresponding to the states at the dot and in the left electrode and
the states in both the electrodes respectively, the current can be
expressed as
\begin{eqnarray}
I_L=\langle
J_L(t)\rangle=&-&\frac{2e}{h}Re[\sum_{k,\sigma}V_LG^<_{dLk\sigma}(t,t)\nonumber\\
&+&\sum_{k'k\sigma} T_{LR}G^<_{CLk\sigma,CRk'\sigma}(t,t)].
\end{eqnarray}
According to the Langreth's rule and using steady-state condition
$I=\frac{I_L-I_R}{2}$, the current can be expressed as
\begin{equation}
I=\frac{2e}{h}\Gamma\int
d\omega\rho(\omega)[f_R(\omega)-f_L(\omega)]ImG_{d\sigma}(\omega),
\end{equation}
where $\Gamma=\pi(|V_L|^{2}+|T_{LR}|^{2})$. For the zero bias
voltage we find the conductance
\begin{equation}
G=-\frac{2e^2}{h}\Gamma\int d\omega\rho(\omega)\frac{\beta
e^{\beta(\omega-\varepsilon_F)}}{[e^{\beta(\omega-\varepsilon_F)}+1]^2}ImG_{d\sigma}(\omega).
\end{equation}
At finite temperatures the zero bias conductance $G$ can be
calculated numerically through the Green function
$G_{d\sigma}(\omega)$ of the QD. The relative position of the level
$\Delta\varepsilon=\varepsilon_F-\varepsilon_d$ can be varied by the
gate voltage applied to the QD.  The temperature $T$ is taken to be
$10^{-5}\Delta$ lower than the Kondo temperature. The hopping matrix
element $V_{a}$ is taken to be 0.1 and the magnetic flux $\Phi$
enclosed in the ring is taken to be $0.5hc/e$. Figure $2$ presents
the zero bias conductance $G$ as a function of $\Delta\varepsilon$
for $T_{LR}=0.01$. The position of maximum conductance is slightly
away from $\Delta\varepsilon=0$, which comes from the Kondo effect.
The asymmetry line-shape comes from the Fano effect. By employing
the cluster expansions, the EOM of Green's functions are transformed
into the corresponding EOM of connected Green's functions. With the
method under the Lacroix approximation, we have calculated the DOS
of the QD and the zero conductance of the system, in which the Kondo
resonance and the Fano effect have been shown. It indicates our
numerical method is reasonable.
\begin{figure}[tbp]
\includegraphics[width=0.51\textwidth]{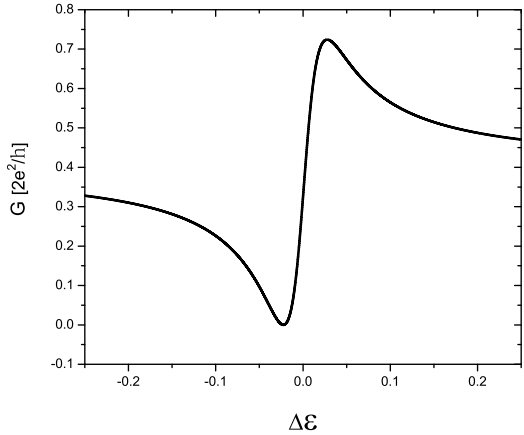}
\caption{Zero bias conductance as a function of $\Delta\varepsilon$
for $T_{LR}=0.01$. The asymmetry line-shape comes from Fano effect.}
\end{figure}

In a similar way we calculate the source-drain voltage properties of
the device. It was assumed that the potential $V$ is applied to the
left electrode and at the right electrode the potential is kept
zero. The relative position of the level
$\Delta\varepsilon=\varepsilon_F-\varepsilon_d$ is hold at 0.04,
which lies in the Kondo regime. The hopping matrix element $V_{a}$
is taken to be $0.1$ and the magnetic flux $\Phi$ enclosed in the
ring is taken to be $0.5hc/e$. The temperature $T$ is taken to be
$10^{-5}\Delta$ lower than the Kondo temperature. Figure 3 presents
the differential conductance $dI/dV$ as a function of $V$ at the
different direct tunneling matrix elements $T_{LR}$. The case for
the pure quantum dot ($T_{LR}=0$) is shown by solid line. Because
the relative position $\Delta\varepsilon$ of the level of the
quantum dot lies in the Kondo regime, the differential conductance
curve shows a very narrow peak at low voltage, which is just the
Kondo resonance observed experimentally as reported in Ref.[14, 15].
The broad maximum seen in Fig. 3 comes from the Lorentzian
\begin{figure}[tbp]
\includegraphics[width=0.5\textwidth]{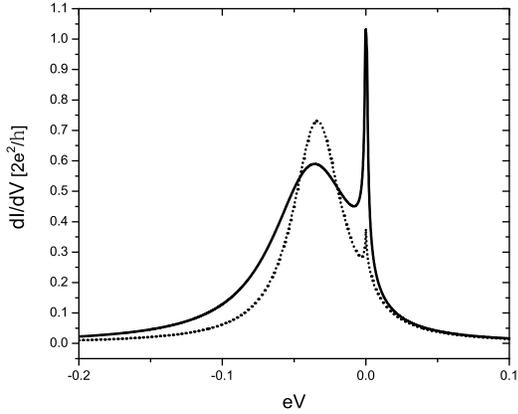}
\caption{Differential conductance for $T_{LR}=0$ (solid line) and
$T_{LR}=0.01$ (dotted line).}
\end{figure}
resonance tunneling when the chemical potential $\varepsilon_F+eV$
approaches $\varepsilon_d$. The influence of the direct channel is
shown by dotted line. The direct electron transmission
($T_{LR}=0.01$) enhances the differential conductance but suppresses
the Kondo resonance peak. It shows the existence of the direct
channel induces the decoherence of the Kondo singlet. However this
is only a phenomenon. Its essence lies in the phase-sensitive
detection of the ABI because the states of the direct channel
decides whether the ABI possesses a phase-sensitive detection
function or not. When the direct channel is not zero, even though
the phase of the QD is not measured, the ABI possesses a
\emph{potential} phase-sensitive detection function and a strong
dephasing of the Kondo singlet is induced. The results of Fig. 3
provide a qualitative explanation with the anomalous features
observed in a recent experiment by Avinun-Kalish \emph{et al}.
[Phys. Rev. Lett. \textbf{92}, 156801 (2004)] and indicates that the
theory \cite{5} of K. Kang is incomplete and the depasing effect of
the phase-sensitive detection of the ABI should also be taken
account.

In conclusion, by employing the cluster expansions, the EOM of
Green's functions are transformed into the corresponding EOM of
connected Green's functions. With the method under the Lacroix
approximation, we have calculated the DOS of the QD and the zero
bias conductance of the system, in which the Kondo resonance and the
Fano effect are shown. It indicates our numerical method is
reasonable. In a similar way we have calculated the differential
conductance and shown that the Kondo assisted transport is
suppressed by the phase-sensitive detection of the ABI. Our
numerical results have provided a qualitative explanation about the
anomalous features observed in a recent dephasing experiment by
Avinun-Kalish \emph{et al}. We have also pointed out that the theory
of K. Kang is incomplete and the dephasing effect due to the
phase-sensitive detection of the ABI should also be taken account.

\end{document}